\documentclass[11pt]{aa}
\usepackage[english]{babel}
\usepackage{graphicx}

\onecolumn
\newcommand{\arep}{Astron. Rep. }

\newcommand{\araa}{Ann. Rev. Astron. Astrophys. }
\newcommand{\mnras}{Mon. Not. R. Astron. Soc. }
\newcommand{\apj}{Astrophys. J. }

\newcommand{\aj}{Astron. J. }
\newcommand{\aap}{Astron and Astrophys.}
\newcommand{\aaps}{Astron and Astrophys. Suppl.}
\newcommand{\pasp}{Publ. Astron. Soc. Pasif. }

\begin{document}

\title{Hypergiant V1302\,Aql in 2001--2014}
\author{V.G. Klochkova,$^{1}$\thanks{E-mail: valenta@sao.ru}, E.L. Chentsov,$^{1}$ A.S. Miroshnichenko,$^{2}$ V.E. Panchuk,$^{1}$  and M.V. Yushkin$^{1}$}
\institute{$^{1}$ -- Special Astrophysical Observatory RAS, Nizhnij Arkhyz,  369167 Russia \newline
           $^{2}$ -- Department of Physics and Astronomy, University of North Carolina at Greensboro, Greensboro, NC 27402--6170, U.S.A.}

\date{\today}
\titlerunning{\it Hypergiant V1302\,Aql in 2001--2014}

\authorrunning{\it Klochkova et al.}

\date{\today}

\abstract
{We present the results of a study of spectral features and the velocity field in the atmosphere and
the circumstellar envelope of the hypergiant V1302\,Aql, the optical counterpart of the IR source
IRC+10420, based on high-resolution spectroscopic observations obtained in 2001--2014.
We measured radial velocities of the following types of lines: forbidden and permitted pure emissions,
absorption and emission components of lines of ions, pure absorptions (e.g., He\,{\sc i}, Si\,{\sc ii}),
and interstellar components of the Na\,{\sc i} D--lines, K\,{\sc i}, and DIBs. The heliocentric radial velocity
measured for pure absorptions as well as for the forbidden and permitted pure emissions is close to the
systemic radial velocity and equal to V$_{\rm r}$\,=\,63.7$\pm$0.3, 65.2$\pm$0.3, and 62.0$\pm$0.4 km\,s$^{-1}$,
respectively. Positions of the absorption components of the lines with inverse P\,Cyg profiles are stable
and indicate the presence of clumps moving toward the star with a velocity of $\sim$20 km\,s$^{-1}$.
The average radial velocity of the DIBs is V$_{\rm r}$(DIB)\,=\,4.6$\pm$0.2 km\,s$^{-1}$.
Comparison of the absorption lines observed in 2001--2014 and those in earlier data shows no noticeable
variations. We conclude that the hypergiant reached a phase of slowing down (or termination) of the effective
temperature growth and is currently located near the high-temperature boundary of the Yellow Void in the
Hertszprung-Russell diagramme.
}

\maketitle

\section{Introduction}\label{introduction}

The evolutionary status of a very luminous star V1302\,Aql, the optical counterpart of the IR source IRC+10420,
has been unclear. The variety of its properties allowed to classify it either as a Proto-Planetary Nebulae (PPN)~[\cite{hkv89}] 
or as a very massive star that has passed through the red supergiant phase~[\cite{jhg93}]. PPNe are currently thought to be low-mass peculiar
supergiants with strong IR excesses in a short-term transition from the Asymptotic Giant Branch to the Planetary Nebula (PN) stage.
They are descendants of intermediate-mass stars (initial masses 1--8 M$_{\odot}$) which have passed several evolutionary
stages, including switching energy sources and stages with typical mass loss rates up to 10$^{-5}$\,M$_{\odot}$\,yr$^{-1}$ and
even up to 10$^{-4}$\,M$_{\odot}$\,yr$^{-1}$. As a result, a PPN is a low-mass degenerate C--O core surrounded by a tenuous and
usually an asymmetric envelope. As the core contracts, its effective temperature (T$_{\rm eff}$) rises and the star moves 
blueward in the Hertsprung-Russell diagramme (hereafter HRD). This stage may last until T$_{\rm eff}$ reaches $\sim$30000 K,
when ionisation of the circumstellar envelope begins. At this time, the object is observed as a PN after hydrogen recombination
lines and forbidden lines of light elements show up in its spectrum. Luminosity of low-mass supergiants at the PPN stage may
reach $\log$~L/L$_{\odot} \sim$ 4.5~[\cite{blocker95}].

Yellow supergiants are significantly different from PPNe, although a number of properties (high luminosity, spectral features,
presence of circumstellar envelopes) are similar in these two classes of object. Predecessors of the former are massive
(initial mass $\ge 20$ M$_{\odot}$) and the most luminous stars, which lose a significant part of their mass after leaving
main-sequence, become red supergiants, and later proceed to yellow supergiants. A typical luminosity of a yellow
supergiant is $\log$ L/L$_{\odot} \sim$ 5.3--5.9~[\cite{jhg93}]. These objects are located near the Eddington limit in an instability
region which contains hypergiants of spectral types from A to M~[\cite{dejager98,dejager01}]. Wolf-Rayet stars and Luminous Blue Variables
may be their descendants~[\cite{oudm09}]. Structured circumstellar envelopes of hypergiants formed during several
phases of a strong stellar wind with mass loss rates of $10^{-4}-10^{-3}$ M$_{\odot}$\,yr$^{-1}$ are sources of IR and maser
radiation as well as of numerous molecular emission lines. Nevertheless, the yellow hypergiant $\rho$~Cas, nearest to V1302\,Aql
in HRD and possessing an extended and unstable atmosphere [\cite{klochkova14}], shows no signs of circumstellar material~[\cite{shm06}].

Obviously, depending on the adopted nature and hence luminosity of an object, its distance estimate may differ by a factor of a few.
However, data obtained during the last two decades from various observations leave no doubts that V1302\,Aql is an object at the
yellow hypergiant stage (see review by Oudmaijer {\it et al.}~[\cite{oudm09}]) as well as later papers~[\cite{driebe09,odw13}].
Moreover, V1302\,Aql is now considered to be the most unambiguous massive Galactic object with a highest mass loss rate which 
undergoes a short-term evolutionary transition from a red supergiant to a Wolf-Rayet star~[\cite{mm03}]. One of the most compelling 
arguments confirming its high-luminosity massive star
status has been derived from spectroscopic data obtained at the 6\,m telescope of the Russian Academy of Sciences, when the
authors~[\cite{kcp97}] found a significant nitrogen excess in the atmosphere of V1302\,Aql. In all the spectra obtained between 1997
and 2000, a He\,{\sc i}\,5876\,\AA\ lines with a large equivalent width of $\ge$200\,m\AA{} was detected. With the
object's T$_{\rm eff} \sim$9200 K, such an equivalent width can be a consequence of a high luminosity and an enhanced
helium abundance in its atmosphere~[\cite{mir13,klochkova02}].

Interest to V1302\,Aql has been growing in the last decade due to detection of a 120\,K a year T$_{\rm eff}$ increase
[\cite{kcp97,klochkova02,oudm96,oudm98}] that allowed to suggest evolution toward the Wolf-Rayet stage.
Humphreys {\it et al.}~[\cite{hds02}] points out that V1302\,Aql is a post red supergiant star which is crossing a critical 
HRD region called the Yellow Void~[\cite{dejager97}].

The T$_{\rm eff}$ increase stimulates us to continue spectroscopic monitoring of this mysterious object. High-resolution
spectroscopy is required to refine the structure and kinematics of its circumstellar envelope. The most adequate model to
represent the observed kinematics is, in our opinion, the ``rain'' model proposed by Humphreys {\it  et al.}~[\cite{hds02}]
even considering a bipolar model suggested later~[\cite{odw13}]. Observations with a moderate spectral resolving power, 
$R \sim 8000$, were obtained by Humphreys {\it et al.}~[\cite{hds02}] to test the model. We obtained spectra with a much higher resolution (up 
to $R \sim 60000$) that allows us to better constrain the line profiles and measure their parameters with a much higher accuracy.

In this paper, we present the results of a new stage of a spectroscopic monitoring of V1302\,Aql in continuation of our
study of this object~[\cite{kcp97,klochkova02}]. In Section~\ref{observations} we describe our spectroscopic data, Section~\ref{discussion}
is devoted to discussion of the results, while Section~\ref{conclusions} summarizes the conclusions.

\begin{table*}
 \centering
  \caption{Averaged heliocentric velocities for groups of lines in the spectra of V1302\,Aql.}
\begin{tabular}{ccccccccc}  
\hline
Date & $\Delta\lambda$,   &\multicolumn{7}{c}{\small V$_{\rm r}$, km\,s$^{-1}$} \\
\cline{3-9}
         &    nm  & \multicolumn{2}{c|}{\small Emissions} &  Em/Abs  &\multicolumn{2}{c|}{\small Absorptions} &\multicolumn{2}{c}{\small IS}\\
         &            &  perm. &   forb.    &   Fe\,{\sc ii}, etc. & Si\,{\sc ii}+He\,{\sc i} & H$\alpha$+H$\beta$  & Na\,{\sc I} D   & DIB \\  
\hline 
09.08.01   &  510--670  &   65  &   68   &   36/79   &  61    & 68 & --   & 5.6  \\ 
28.08.04  &  530--680  &   67  &   61   &   43/83   &  68    & 69 & 11.3 & 4.8  \\   
24.11.07  &  530--680  &   65  &   61   &   42/77   &  58    & 68 & 11.2 & 4.5  \\
13.07.08  &  520--670  &   65  &   65   &   42/79   &  54    & 67 & 11.2 & 5:   \\
18.08.08  &  460--600  &   66  &   65   &   45/80   &  60    & 70 & 11.0 & 5:   \\
3 \& 5.11.08 &  450--590  &   64  &   60   &   46/85   &  71    & 72 & 11.0 & 4.4  \\
11.09.09  &  420--880  &   64  &   60   &   39/82   &  66    & 69 & 10.6 & 3:   \\
31.07.10  &  440--590  &   66  &   59   &   42/80   &  60    & 74 & 10.8 & 5:   \\
20.11.10  &  400--550  &   64  &   --   &   43/80   &  --    & 70 & --   & --   \\
3 \& 8.08.12& 430--680  &   65  &   62   &   37/81   &  70    & 72 & 10.6 & 4.4  \\
27.05.13  &  430--670  &   64  &   61   &   40/81   &  65    & 73 & 10.5 & 4.8  \\
19.08.13  &  430--670  &   66  &   61   &   35/80   &  64    & 72 & 10.3 & 5.2  \\
09.10.13  &  430--670  &   66  &   61   &   37/82   &  67    & 72 & 10.8 & 4.6  \\     
13.08.14  &  430--670  &   66  &   62   &   38/83   &  63    & 72 & 11.1 & 4.2  \\
04.10.14   &  540--850  &   65  &   62   &    --     &  65    & 70 & 10.5 & 4:   \\
\hline    
\multicolumn{2}{c}{}  & \multicolumn{7}{l}{\small\underline{Average values of V$_{\rm r}$, km\,s$^{-1}$}} \\ 
 \multicolumn{2}{c}{} & 65.2   & 62.0   & 40.5/80.7   &  63.7  &70.5      &  10.8    & 4.6 \\
 \multicolumn{2}{c}{} &$\pm$0.3&$\pm$0.4& $\pm0.5/0.4$&$\pm0.3$& $\pm$0.4 & $\pm$0.2 & $\pm$0.2 \\    
\hline
\end{tabular}
\end{table*}

\section{Observational Data}\label{observations}

We have added 16 new spectra to our collection of the V1302\,Aql data in 2001--2014. The observing dates and data spectral ranges are
listed in the first two columns of Table\,1. Most spectra were obtained with the \'echelle spectrograph NES with spectral resolution 
$R = 60000$~[\cite{pyn03,panchuk09}]. Our first spectrum obtained on 2001 August 9 was taken with the spectrograph
LYNX 1K$\times$1K CCD, $R \sim 25000$~[\cite{panchuk93}]. One-dimensional data were extracted from 2D \'echelle spectra
using the {\it ECHELLE} context in {\it MIDAS} modified to features of the spectrographs used (see details in~[\cite{yk05}]).
Cosmic particles were removed by median averaging of two consecutive spectra. Wavelength calibration of the
spectra was derived using a Th-Ar hollow cathode lamp.

One of the spectra used in this paper was obtained at the McDonald Observatory on 2009 September 11 with the
\'echelle spectrograph TS2 $R = 60000$ [\cite{tull95}] in the coud\'e focus of the 2.7\,m Harlan J.\,Smith telescope.
This allowed us to fill the gap in observations with the 6\,m telescope
for 2009. One-dimensional data for this spectrum were extracted using the {\it apall} task within the {\it echelle}
package in IRAF. In particular, the spectral range of this observation allowed us to study profiles of the strong lines of
Ca\,{\sc ii} 8408\,\AA\ and 8542\,\AA\ in the near IR region, which is not covered by the 6\,m telescope data.

The remaining part of the data reduction, including measurements of the lines intensities and positions, was done with the latest
version of the DECH20t package~[\cite{g92}]. This package, traditionally used in our studies, permits radial velocity
measurements for individual features of complex line profiles. Only heliocentric radial velocities, V$_{\rm r}$, are used
throughout this paper. Their systematic errors do not exceed 1\,km\,s$^{-1}$ for a single line. The latter can be seen in
the last column of Table\,1, where V$_{\rm r}$ of the saturated interstellar components of the Na\,{\sc i} D--lines are
listed. Features in the spectrum of V1302\,Aql were identified using an earlier published atlas~[\cite{ckt99}].

\section{Results and Discussion}\label{discussion}

\subsection{Spectral type of V1302\,Aql in 2001--2014} 

Earlier spectroscopic observations of V1302\,Aql revealed a gradual transition from a normal F--type
supergiant~[\cite{humph73}] to an A5--type one~[\cite{klochkova02}]. The new spectral type estimates for 2001--2014
based on our technique presented in [\cite{kcp97,klochkova02}] give the following results. Pure absorption lines in
the blue part of the spectrum indicate a spectral type A6, while inclusion of criteria using He\,{\sc i} and
Si\,{\sc ii} absorptions leads to A3.5. Therefore we conclude that the current spectral type coincides with
that derived in [\cite{kcp97,klochkova02}] within the uncertainties and that the objects T$_{\rm eff}$ does not grow anymore.

\subsection{Line profiles in the spectrum of V1302\,Aql in 2001--2014} 

\begin{figure}     
\includegraphics[angle=0,width=0.4\textwidth,bb=30 30 360 740,clip=]{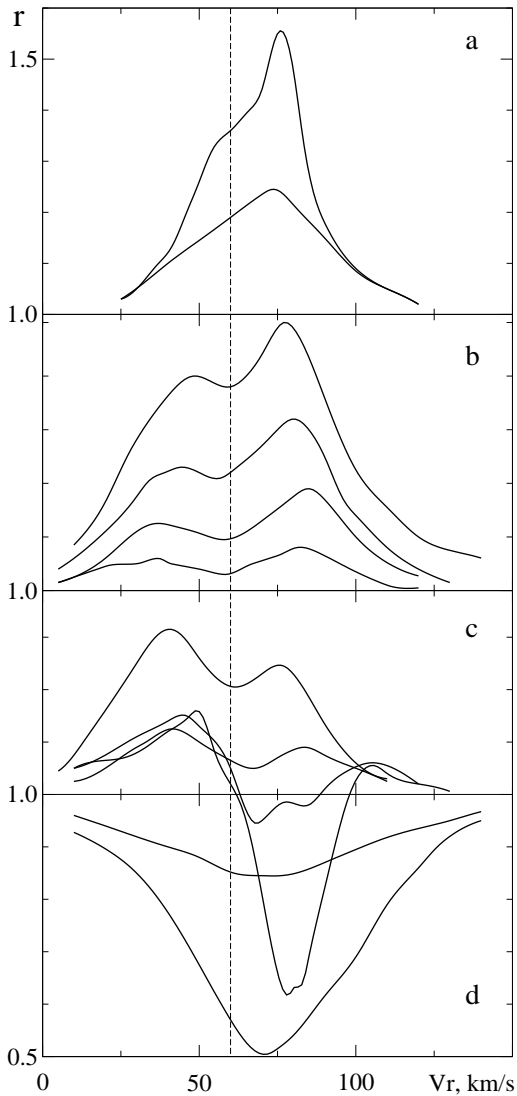}
\caption{The line profile transformation in the spectra of V1302\,Aql in 2012--2014. From top to bottom:
a) the forbidden emissions [Fe\,{\sc ii}] (Multiplet 14F) 7155\,\AA\ and [O\,{\sc i}] (1F) 6300\,\AA ; 
b) the emissions Fe\,{\sc ii} (46) 5991, 6084, and 6113\,\AA\ and an average of the Fe\,{\sc i} (168) 6394\,\AA\
and Ti\,{\sc ii} (112) 6718\,\AA{};  c) the emission Fe\,{\sc ii} (74) 6417\,\AA\, an average of the emissions 
Cr\,{\sc ii} (50) 5502\,\AA\ and 5511\,\AA\, and absorption/emission lines Ti\,{\sc ii} (69) 5337\,\AA\ and Ti\,{\sc ii} (70) 5154\,\AA\
d) the absorptions Si\,{\sc ii} (5) 5056\,\AA\ and Si\,{\sc ii} (2) 6347\,\AA{}. The vertical line shows the systemic
velocity V$_{\rm sys} \sim 60$\,km\,s$^{-1}$ [\cite{oudm96}].}
\label{fig1}
\end{figure}

As earlier, line profiles in the spectra of V1302\,Aql vary from purely absorption with small deviations from symmetry
to P~Cyg--type profiles and double-peaked emissions. Examples of these line profiles are shown in Fig.\,\ref{fig1}.
Variations of the intensities and profiles for the last 20 years taking into account results from [\cite{kcp97,klochkova02,hds02}] 
are noticeable but small.

\begin{figure}
\includegraphics[angle=-90,width=0.5\textwidth,bb=10 30 590 780,clip=]{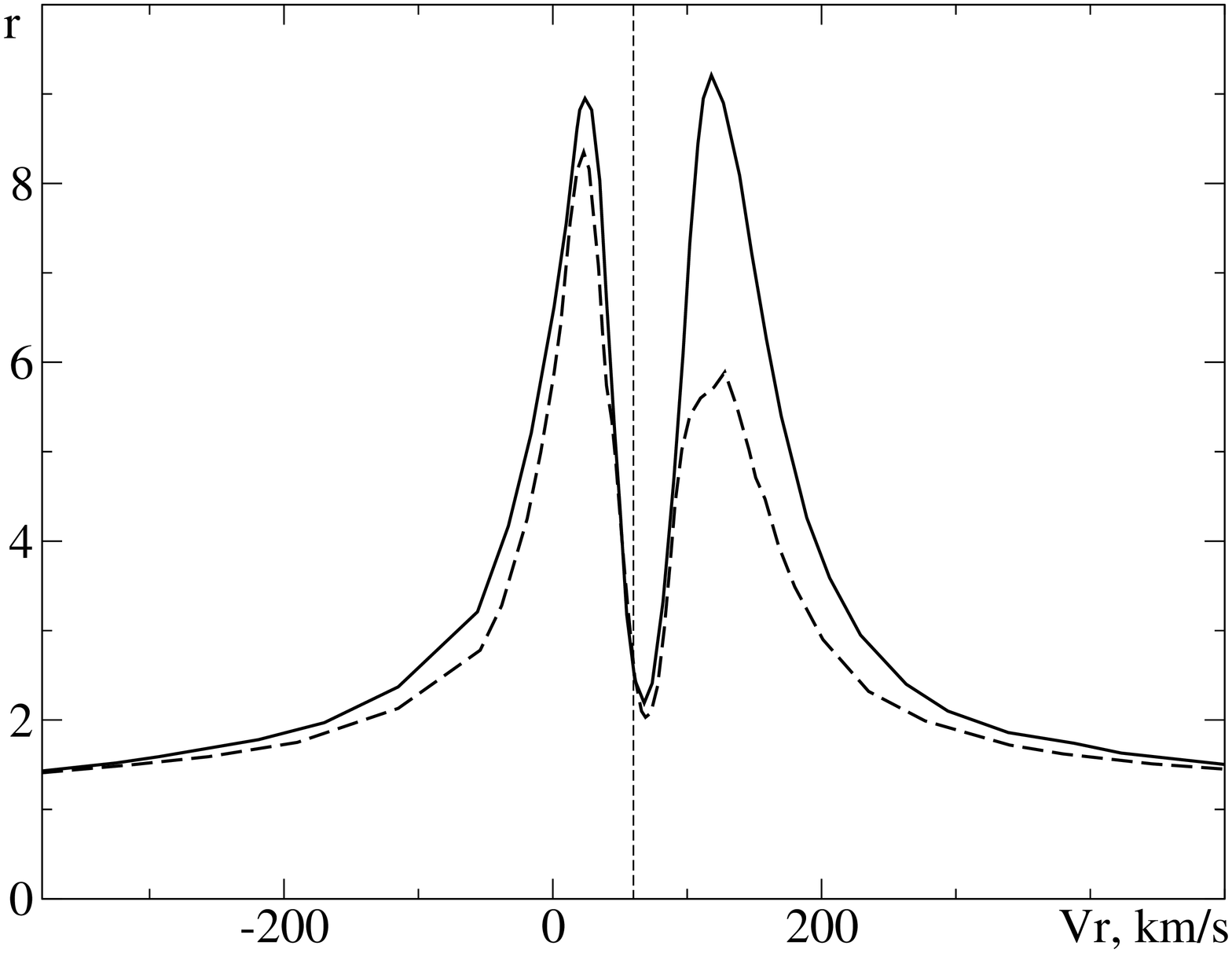}
\caption{The H$\alpha$ line profile in the spectra of V1302\,Aql in 2007 and 2014. The vertical  line shows the systemic
      velocity V$_{\rm sys} \sim 60$ km\,s$^{-1}$ [\cite{oudm96}].}
\label{fig2}
\end{figure}

Hydrogen lines H$\alpha$ and H$\beta$ in 2001--2014 still have a characteristic double-peaked profile, which was
observed in the 1990's see Fig.\,5 in~[\cite{klochkova02}]. Positions of both emission peaks and the central depression have not
changed since then. Variations of the residual intensities of the emission components of both lines remained within 40\%.
The H$\alpha$ profiles in two spectra are shown in Fig.\,\ref{fig2}.

A typical peak intensity ratio for the entire 20-year long period is seen in the spectrum taken in 2014.
An unusual H$\alpha$ line profile, when the red peak is significantly stronger than the blue one, is detected in only one
spectrum taken on 2007 November 24. In all other spectra the peak intensity ratio is the opposite [\cite{klochkova02, hds02}].
Variations of the residual intensities of the strongest forbidden and permitted Fe\,{\sc ii} lines are even weaker (within 10\%).
In the last 7 spectra taken in 2012--2014 they are limited to 20\% and 6\%, respectively.

\subsection{Radial velocities of various features in 2001--2014} 

According to~[\cite{oudm96}], the average radial velocity of several rotational bands of the CO molecule with respect to the local
standard of rest is V(LSR)\,=\,77\,km\,s$^{-1}$. The heliocentric systemic radial velocity of the object is V$_{\rm sys} \sim$
60\,km\,s$^{-1}$.  Humphreys {\it et al.}~[\cite{hds02}] determined V$_{\rm sys} = 58-60$ km\,s$^{-1}$ using a combination of CO and OH bands.
We note that the latter is also close to the velocities derived from radio lines of other molecules
(see~[\cite{ql13}] and references therein).

\begin{figure}
\includegraphics[angle=-90,width=1.0\textwidth,bb=0 0 93 320,clip=]{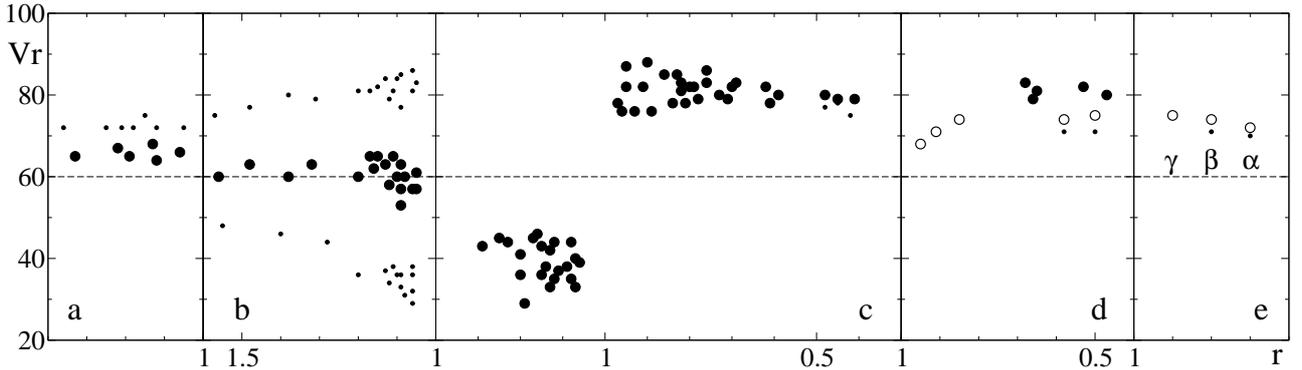}
\caption{The relationships between the heliocentric radial velocities and residual intensities of the line components
    in the spectra of V1302\,Aql on 2008 August 3 and 8. The horizontal dashed line shows the systemic
    velocity V$_{\rm sys} \sim 60$\,km\,s$^{-1}$ [\cite{oudm96}].
    a) Forbidden emissions (whole profiles are shown by filled circles, while their peaks are shown by dots).
    b) Permitted double-peaked profiles (symbols are the same as in part a).
    c) Lines with inverse P~Cyg--profiles (emission and absorption components are shown by filled circles, while cores of the
    strongest absorptions are shown by dots).
    d) Fe\,{\sc ii} absorptions (whole profiles are shown by circles) and He\,{\sc i} and Si\,{\sc ii} absorptions (whole profiles
    are shown by open circles, cores are shown by dots).
    e) Absorption components of the H\,{\sc i} lines (symbols are the same as in part d). }
\label{fig3}
\end{figure}

The relative stability of the spectrum of V1302\,Aql in recent time and homogeneity of the obtained material allowed us
to move from reporting the variety of the spectral line shapes to following a gradual transformation between different
shapes. Comparison of Figs.\,\ref{fig1} and \ref{fig3} with the data from Table 1 convinces that such a transformation is real.
The profiles shown in Fig.\,\ref{fig1} are averaged from the spectra taken in 2012--2014. A typical example of relationships
between the radial velocities for individual lines and their residual intensities is shown in Fig.\,\ref{fig3}. The same profile types
follow each other from top to bottom in Fig.\,\ref{fig1} and from left to right in Fig.\,\ref{fig3} and in Table\,1.

Variations of the average radial velocities of various line groups (see Table\,1) are small as well. In particular, all the
radial velocities for the pure absorptions are in a range of 54--70\,km\,s$^{-1}$. Forbidden lines are asymmetric
(see Fig.\,\ref{fig1}a), such as the peak intensity is red-shifted with respect to the lower part of the profile by
6\,km\,s$^{-1}$ on average. The [Ca\,{\sc ii}](1F) 7291 and 7324\,\AA\ emissions seen in our spectra taken on
2009 September 11 and 2014 October 4 are much stronger that all other forbidden lines but with no positional
shift with respect to the latter. Radial velocities of the entire profiles and their peaks are presented in Fig.\,\ref{fig3}a
by circles and dots, respectively. The averaged radial velocities, measured from lower parts of the forbidden
line profiles, are listed in column 3 of Table~1.

Permitted emissions of the iron group are noticeably wider than the forbidden lines. For example, at the same residual
intensity of 1.5, the half-widths at the zero level are 70 and 50\,km\,s$^{-1}$, respectively. The iron lines are clearly
double-peaked with a weaker blue-shifted peak (see~Fig.\,\ref{fig1}b). The radial velocities of the lower parts of these
emission profiles (in the same way as for the forbidden lines), are shown in column 4 of
Table~1 and represented by circles in Fig.\,\ref{fig3}b. The emission peak velocities are shown by dots in the same Figure.
The peak separation decreases from 50\,km\,s$^{-1}$ for weak emissions to 26 km\,s$^{-1}$ for the strongest ones,
where it even reaches a triangular shape. It seems that the double-peaked profiles are not just a sum of two narrower
single-peaked emissions separated by 50\,km\,s$^{-1}$, but that a partially filled absorption component formed in the stellar
atmosphere also takes part in the overall profile formation. 

It is seen in Figs.\,\ref{fig1}bc that a gradually deepening absorption ``pushes down''
the central part of the profile and eventually turns it into an inverse P~Cyg--type profile. Line profiles may have this kind of
shape in the following cases: when a narrow circumstellar absorption overlaps with a wide circumstellar emission or when two
narrower emissions with different radial velocities join together. The inner slopes would be shallower than the outer ones
in the latter case. We observe steeper inner slopes instead, therefore the emission components are separated by an absorption.
It is possible to have a combination of these two formation mechanisms as well as a double--ray version suggested in~[\cite{odw13}].

There a many lines with emission components on both sides of the absorption in the spectrum of V1302\,Aql. We include
in the group of lines with an inverse P~Cyg--profile those with the continuum above the absorption core, $r <$1.
If the central depression is seen at $r >$1, the line is considered to have a double-peaked profile. Column 5 of Table~1
contains the average radial velocities for blue-shifted emission and absorption components in the group of lines with
P~Cyg--profiles. Both such components are shown by circles in Fig.\,\ref{fig3}c (an absorption asymmetry is shown for
the strongest ones). The red-shifted components are weaker ($r <$1.1) than the blue-shifted ones with an average radial
velocity of 130$\pm$10\,km\,s$^{-1}$ in our spectra.

The profiles in Fig.\,\ref{fig1} differ from classic P~Cyg--profiles which represent a spherically-symmetric envelope with a radial
velocity gradient. The latter show a relationship between the emission component intensity and the absorption component depth.
However the weakest absorptions in our profiles ($r < $0.95) may be accompanied by blue-shifted emissions with any intensity in a
range 1.0 $< r <$ 1.4, while the strongest absorptions ($r \sim$0.4) are seen with the weakest ones ($r <$0.15).
This is seen in Fig.\,\ref{fig3} where the leftmost filled circles of the lower chain correspond to the leftmost (not the rightmost!)
circles of the upper chain.

The line profiles that are totally located below the continuum make the group of absorptions. However, some of these lines,
such as Fe\,{\sc ii} (37,38), Ti\,{\sc ii} (31), Cr\,{\sc ii}, may have hidden outside emission components, because their wings are
narrower than those of Si\,{\sc ii} and Mg\,{\sc ii} with similar depths. Fig.\,\ref{fig3}d shows the uncertain absorption group members
by filled circles, while more reliable ones are shown by open circles. In turn, we consider the most pure absorptions are the weakest
ones, such as Si\,{\sc ii} (4,5) and He\,{\sc i} 5876\,\AA\ (the left subgroup of the open circles in Fig.\,\ref{fig3}d). The average radial
velocities for the latter lines are given in column 6 of Table~1. A standard deviation for the pure absorption and emissions
is $\sigma \le$ 0.4 km\,s$^{-1}$ and $\sigma =$ 0.2 km\,s$^{-1}$ for the interstellar features.

Closeness of the radial velocities seen in Fig.\,\ref{fig3} is noticeable in the following line groups:
\begin{itemize}
\item iron absorption lines in the blue part of the spectrum and absorption components of the lines with inverse
P~Cyg--profiles (averaging all our data gives 81 and 80 km\,s$^{-1}$, respectively). We consider
stable positions of the latter as an indicator of a matter infall;
\item strong Si\,{\sc ii}~(2) absorptions (they are shown by a pair of open circles in Fig.\,\ref{fig3}d) and absorption
components of the H$\alpha$ and H$\beta$ lines (average radial velocities are 73 and $\sim$70\,km\,s$^{-1}$,
respectively);
\item the weakest absorptions and forbidden emissions (63.7 and 65.2 km\,s$^{-1}$, respectively).
Since the former form in the deepest layers of the photosphere (or pseudo-photosphere) while the latter
form in an extended envelope, it is natural that the radial velocity of the latter is the closest to that of the star's
center of mass, i.e. V$_{\rm sys} \sim$ 60\,km\,s$^{-1}$.
\end{itemize}

\section{Interstellar features in the spectrum of V1302\,Aql}

\begin{figure}
\includegraphics[angle=-90,width=0.5\textwidth,bb=15 15 590 780,clip=]{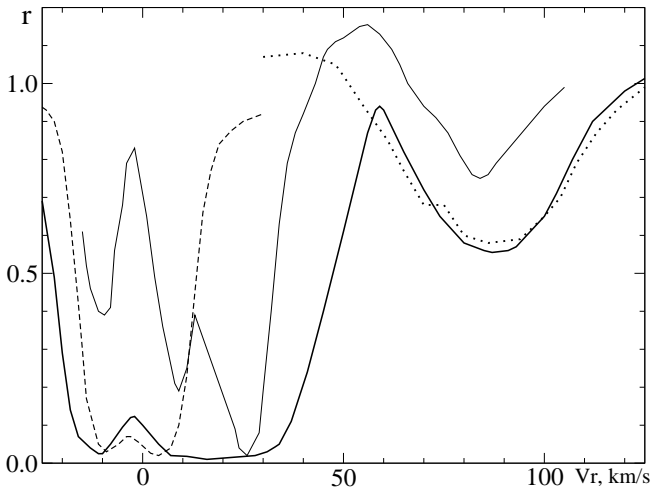}
    \caption{The Na\,{\sc i} D1 line profile in the spectra of V1302\,Aql (thick solid line) and HD\,183143 (dashed line).
    The thin solid line shows the profile of the K {\sc i} (1) 7665\,\AA\ line in the spectrum of V1302\,Aql. The dotted line
    shows the Fe\,{\sc ii} 5316\,\AA\ line profile in the spectrum of V1302\,Aql. Spectra of both stars were taken with the
    same spectrograph NES.}
    \label{fig4}
\end{figure}

Fig.\,\ref{fig4} shows a complex profile of the D1--line of a resonance Na\,{\sc i} doublet. The mail part of its
absorption component forms in a cold interstellar gas in the line of sight, while a weaker red-shifted component
forms in the atmosphere of V1302\,Aql. The latter is well reproduced by the absorption component of the Fe\,{\sc ii}
5316 \AA\ line, and both can apparently be classified as lines with inverse P~Cyg--profiles.

The main part of the Na\,{\sc i} D--line profile in a range of V$_{\rm r} = -25 - +50$\,km\,s$^{-1}$ has interstellar nature.
As follows from a recent compilation of the data on the structure and kinematics of the Milky Way~[\cite{reid14}],
radial velocity in the direction of V1302\,Aql increases with distance and reaches +50\,km\,s$^{-1}$ at a distance D\,=\,5.3\,kpc.
Fig.\,\ref{fig4} shows the Na\,{\sc i} D1--line profile in the spectrum of the B--type hypergiant HD\,183143, which is
projectionally close to V1302\,Aql (the Galactic coordinates $l/b = 53\fdg2/0\fdg6$ and 47$\fdg1/-2\fdg5$, respectively).
However, as seen in Fig.\,1 in~[\cite{ql13}], HD\,183143 is located between the local and Carina--Norma spiral arms,
while the line of sight to V1302\,Aql passes through the latter arm between 3 and 8\,kpc. Therefore, a distance toward
HD\,183143 is $\sim$2\,kpc~[\cite{ch04}], and V1302\,Aql is located significantly further away. Coincidence of the red boundary of the
interstellar Na\,{\sc i} absorption and the maximum of above mentioned V$_{\rm r}$(D) relationship indicates that V1302\,Aql
cannot be closer than 5.3\,kpc. Its spectroscopic parallax moves it even further away to 6.5--8.0\,kpc.

The K\,{\sc i} 7665\,\AA\ line profile shown in Fig.\,\ref{fig4} also represents a combination of stellar and interstellar components.
The interstellar components are less saturated compared to those of the Na\,{\sc i} D--lines and split into three components.
The weakest of them at V$_{\rm r} \sim -10$\,km\,s$^{-1}$ corresponds to the blue-shifted component of the Na\,{\sc i} D1 line
(Fig.\,\ref{fig4}), while the other two at V$_{\rm r} \sim 8$ and 24\,km\,s$^{-1}$ are merged in the Na\,{\sc i} profile.
The stellar absorption component is located at V$_{\rm r} \sim$ 84 km\,s$^{-1}$. As clearly seen in Fig.\,\ref{fig4}, the K\,{\sc i}
7665\,\AA\ line profile also has an emission component, whose position at V$_{\rm r} \sim$ 84\,km\,s$^{-1}$ indicates that it forms
in the circumstellar envelope.

Using our large high-quality material, we measured positions of some diffuse interstellar band (DIB). A list of those reliably identifiable in
the spectrum of V1302\,Aql was published in~[\cite{ckt99}]. Note two important points concerning the search and measurement of the
DIB positions:
\begin{itemize}
\item the number of DIBs with measured V$_{\rm r}$ varies from one spectrum to another (from 5 to 17 features);
\item the individual DIBs V$_{\rm r}$ differ systematically.
\end{itemize}

The latter was mentioned by Oudmaijer~[\cite{oudm98}], who has identified many DIBs in the object's spectrum. Over 30 of them are narrow and have
equivalent widths of $\ge$20\,m\AA. The author explained a large scatter of the measured V$_{\rm r}$ (from 5 to 40\,km\,s$^{-1}$) by
uncertain DIB wavelengths.

The last column of Table~1 lists average V$_{\rm r}$ for each of the observing dates in 2001--2014 of a small group of the narrowest and
most symmetric DIBs which show the smallest differential shifts. Their standard wavelengths (5796.97, 5849.82, 6195.96, 6376.00,
and 6379.24\,\AA) are taken from~[\cite{wes10}]. After such a selection, the average V$_{\rm r}$ for all the dates came to
4.6$\pm$0.2\,km\,s$^{-1}$.

\section{Closest analogs of the hypergiant V1302\,Aql}

In Sect.\,\ref{introduction} we mentioned $\rho$ Cas, a well-studied yellow hypergiant with a luminosity similar to that of V1302\,Aql.
In particular, optical spectra of $\rho$ Cas and a complex structure of its atmosphere and envelope have been recently studied in detail
[\cite{klochkova14,gor06}]. However if all features are taken into account, the optical counterpart of the IR-source IRAS\,18357$-$0604 turns out to be
closer to V1302\,Aql. Since this object is very reddened, not much data have been published for it. Recently authors~[\cite{clark14}] 
took a spectrum of IRAS\,18357$-$0604 in the red and IR regions, estimated its distance D\,$\sim$6\,kpc using its systemic velocity, classified it an
early A--type star, and concluded on a similarity of its spectrum to that of V1302\,Aql. The spectrum of IRAS\,18357$-$0604 is dominated
by asymmetric emissions of low-excitation transitions, such as H\,{\sc i}, N\,{\sc i}, Fe\,{\sc i}, Fe\,{\sc ii}, Ti\,{\sc ii}, [Fe\,{\sc ii}], etc.
According to these authors, line profile features of IRAS\,18357$-$0604 indicate an asymmetry of its outflowing envelope.
However unlike in the spectrum of V1302\,Aql, red-shifted components of broad double-peaked profiles of H {\sc i}, Ca {\sc ii},
and N\,{\sc i} in the spectrum of IRAS\,18357$-$0604 are stronger than blue-shifted ones.

A bright star HR\,8752, the optical counterpart of the IR source IRAS\,22579+5640, is also a member of the yellow hypergiants group.
It is located much closer to the Sun compared to V1302\,Aql, and it is easier to study its features. HR\,8752 is listed as a standard of
MK--classification with a spectral type G0\,Ia~[\cite{mr50}]. However comparing high-resolution spectra taken in 1973--1977, authors~[\cite{LL78}]
have detected a growth of its T$_{\rm eff}$. The same authors studied kinematic properties of the star's atmosphere and envelope in
detail and noted signs of a matter infall onto the star with a speed of 30 km\,s$^{-1}$. Later, based on a set of optical spectra of HR\,8752
obtained with various instruments in 1973--2005, Nieuwenhuijzen {\it et al.}~[\cite{nieuw12}] have analysed these data using a homogeneous approach. 
Adding data from even a longer photometric monitoring, they restored temporal variations of the star's fundamental parameters, such as T$_{\rm eff}$,
luminosity, radius, color-index $B-V$, etc. One of their main results is a conclusion about a gradual growth of T$_{\rm eff}$ from
$\log$ T$_{\rm eff}$\,=\,3.65 circa 1900 to $\log$ T$_{\rm eff}$\,=\,3.90 in 2000. Therefore, HR\,8752 is the closest analog of V1302\,Aql
based on all features. At the same time, HR\,8752 seems to have a lower mass compared to that of V1302\,Aql judging on the HRD
position~[\cite{nieuw12}].

Comparison of the spectral line profiles in our spectra of V1302\,Aql in 2001--2014 indicates the absence of a noticeable spectral
variability, thus allowing to conclude that the hypergiant entered a phase of a slow down (or termination) of the T$_{\rm eff}$ growth
and approached the Yellow Void boundary, which is called the White Wall~[\cite{odw13}]. Earlier authors~[\cite{patel08}] suspected 
stabilization of the star's T$_{\rm eff}$ based on a long-term photometric monitoring. New evolutionary loops may follow this episode, 
such as that observed for HR\,8752 on a timescale of 10 years~[\cite{dejager97}]. Therefore, it seems very important to continue monitoring 
of V1302\,Aql.

\section{Conclusions}\label{conclusions}

Using a set of high-resolution spectra of V1302\,Aql obtained in 2001--2014, we measured intensities and positions of various spectral
features that allowed us to analyse the profile behaviour with time as well as the velocity field in various layers of the object's extended
atmosphere and its circumstellar envelope. We concluded on a closeness of the V$_{\rm r}$ for the iron absorption lines in the blue spectral
part and absorption components with inverse P~Cyg--profiles. Positions of the latter features that reflects clumps falling onto the star
with a velocity of $\sim$20\,km\,s$^{-1}$ has been stable for all the observing dates. The position of the strong Si\,{\sc ii}~(2) absorptions and
absorption components of the H$\alpha$ and H$\beta$ lines varied insignificantly around 63.7 and 70.5 km\,s$^{-1}$, respectively.

The V$_{\rm r}$ for the weakest absorptions, which form in the deepest observable photospheric (or pseudo-photospheric) layers, and
for the forbidden emissions, which form in an extended envelope (63.7 and 65.2 km\,s$^{-1}$, respectively). The average V$_{\rm r}$ of
the permitted emissions also weakly deviates from that of the pure absorptions. It is equal to 62.0\,km\,s$^{-1}$ for all the observing dates.
The average V$_{\rm r}$ of the interstellar features is equal to 4.6$\pm$0.2 km\,s$^{-1}$.

Comparison of the features in the spectra of V1302\,Aql in 2001--2014 indicates the absence of a noticeable variability. We conclude that 
the hypergiant entered a phase of a slow down (or termination) of the T$_{\rm eff}$ growth and approached the high-temperature boundary 
of the Yellow Void.

\section*{Acknowledgments}

This study was accomplished with a financial support of the Russian Foundation for Basic Research (RFBR) in the framework of
the project No.14--02--00291\,a. A.M. acknowledges support of his travel to the McDonald Observatory from the Department of Physics
and Astronomy of the University of North Carolina at Greensboro. We have made use of the astronomical data bases SIMBAD and ADS.

\end{document}